# Percolation, fractals and the critical point in a nuclear reactor

V. V. Ryazanov

Institute for Nuclear Research, pr. Nauki, 47 Kiev, Ukraine, e-mail: vryazan19@gmail.com

The trajectories of neutrons in the reactor, the points of their fission of uranium nuclei, the points of neutron absorption, fission chains and chain reactions are considered from the standpoint of fractal geometry and percolation theory. In the study of the stationary critical operating mode of a nuclear reactor, models of Cayley trees and Laplacian fractals are used. This approach allows us to obtain the neutron multiplication equation and an expression for the critical size of the reactor. Models of irreversible growth and various fractal dimensions are also considered as applied to the evolution of neutrons in a reactor. Prospects for the development of the proposed approach to describing reactors, primarily the kinetics and processes of neutron transfer, are indicated.
Classifications list: 05C05, 05C92, 05C07, 82C26, 82C27, 82C31, 82C43, 82B43, 60K50, 60K35.

## 1. Introduction

Fractal models of various kinds of systems make it possible to discover new features of seemingly well-known phenomena. Many physical systems are fractal and multifractal. Below we consider some possibilities for describing the motion of neutrons in a reactor using the approaches of percolation [1, 2] and fractality [2–4]. Percolation is the phenomenon of flow or non-flow of liquids through porous materials, electricity through a mixture of conductive and non-conducting particles, and other similar processes. Percolation theory finds application in describing a variety of systems and phenomena, including the spread of epidemics and the reliability of computer networks. Percolation is the moment when a system state appears in which there is at least one continuous path through adjacent conducting nodes from one to the opposite edge. The set of elements through which flow occurs is called a percolation cluster. In [5], fractal is a structure consisting of parts that are in some sense similar to the whole. Fractal properties manifest themselves especially clearly at the very point of the phase transition, in the critical region. The steady-state operation of a nuclear reactor (NR) takes place precisely at the critical point, and the fractal description should be very important for characterizing the neutron behavior in the reactor [6].

In [7], the spread of rumors in the percolation model is compared with a chain reaction. The relations of the theory of percolation [1, 2] are also valid in the general theory of phase transitions [8 - 10]. Fractal concepts were used in the study of highly developed turbulence, inhomogeneous star clusters [11], diffusion-limited aggregation, processes of destruction of matter, the structure of blood, etc. The description of the physical properties of systems with a fractal structure led to the development of analytical methods in the fractal concept based on the use of the mathematical apparatus of fractional order integrodifferentiation, since the dimension of space becomes fractional. The need to switch to neutron transfer equations in fractional derivatives may be of practical importance for reactor calculations [12], although there is often no sharp distinction between percolation processes and diffusion [2]. It was noted in [27] that transport processes in percolation clusters, fractal trees, and porous systems must be analyzed anew in order to obtain correct transport equations for such systems. In branching fractal structures, "super-slow" transfer processes can occur, when a physical quantity changes more slowly than the first derivative. The index of the fractional derivative with respect to time corresponds to the proportion of channels (branches) open to flow. The dynamics of diffusion is determined by the random nature of particle motion. A diffusing particle can reach any point in the medium. Percolation is associated with a fractal environment. Below the occurrence threshold, the process of particle propagation is limited to a finite region of the medium. During diffusion from a source, a diffusion front appears that has a fractal structure. In [5], the term shell of a percolation cluster is introduced. Below, from a fractal perspective, we consider the chain reaction processes in the reactor: neutron trajectories, fission of uranium nuclei by neutrons, the formation of new neutrons, their absorption, etc.

## 2. Fractal behavior of fission chains. Percolation models

The model of Cayley trees [13 - 15] or Bethe lattice is applicable to fission chains in NR. A Cayley tree, also called a Bethe lattice, is constructed starting from a central node from which z branches of unit length emanate, forming the first shell of the Cayley tree. The end of each branch is also a node. From each node z - 1 new branches emanate, forming z(z - 1) nodes of the second shell. The process continues ad infinitum. This produces an infinite Cayley tree with z branches emanating from each node. There are no loops in the system because any two nodes are connected by only one path. In this case, the random nature of branching should be taken into account. You can also apply the theory of random graphs [16]. In [17], objects are considered that are very close to the subject of our study - random trees (random aggregates without loops). A general relationship between the diffusion index and the fractal dimension of the tree is obtained. Knowledge of the internal properties of clusters allows one to study their dynamic properties. Fractal phenomena can be classified as stochastic phenomena, since there is a close connection between fractal phenomena and statistical distributions [18].

The trajectories of neutron motions in nuclear reactors form a characteristic branching structure of the process for the total number of neutrons. The figure shows examples of the trajectories of one neutron introduced into a breeding medium, taking into account those evolutionary events (nuclear fission and neutron absorption) that lead to a change in the size of the neutron population.

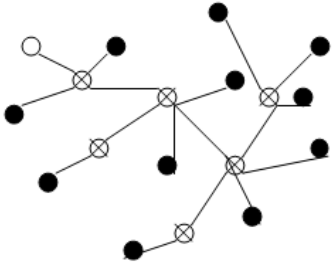

Fig. 1. Trajectories of neutrons and their descendants in the breeding medium: ○ - the point at which the initial neutron begins to move; ⊗ —points of fission of nuclei by neutrons; ● - neutron absorption points.

The processes represented by trees are associated with branching random processes [19]. We consider an aspect of the problem determined by the size and nature of the behavior of clusters - nodes connected to each other. By a node we mean a fissile nucleus (or a neutron introduced into the system - the root of a tree [16]), and by a connection - neutron trajectories. Neutron absorption points form so-called hanging ends [16] (vertices of degree 1) or free ends.

At high subcriticality and large negative reactivity values $\rho=(k_{ef}-1)/k_{ef}$ (the effective multiplication factor $k_{ef}$ is much less than unity), the system contains small-sized clusters with a predominant number of hanging ends. If the intensity of neutron death during the time $\Delta t \to 0$ (absorption by the environment or leaving the system) is denoted $\lambda_c \Delta t + 0(\Delta t)$, and the intensity of nuclear fission by a neutron $\lambda_f \Delta t + 0(\Delta t)$ ($\lambda_f = v\Sigma_f$, where v is the neutron speed; $\Sigma_f$ is macroscopic fission cross section), then the probability of nuclear fission by a neutron is $c=p=\lambda_f/(\lambda_f+\lambda_c)$. Effective neutron multiplication factor $k_{ef} = p\bar{v}$, where $\bar{v}$ is the mathematical expectation of the number of secondary neutrons in one fission event. As *p* increases, the cluster sizes increase. At *p* = 1, all fuel nuclei in the nuclear reactor are separated, and $k_{ef\,max}= \bar{v}$ (under such conditions an explosion occurs). For $1 - p \ll 1$ there is an infinite cluster in the system. There must be a critical value $p_c$ at which a transition from one mode to another occurs - an infinite cluster appears for the first time. This corresponds to the case $k_{ef} = 1$, $p_c = 1/\bar{v}$. The same result in the percolation model was obtained strictly mathematically [1 - 4, 7]. The formation of an infinite cluster represents a phase transition, the beginning of a self-sustaining chain reaction, the critical point of the system. An important role in the theory of phase transitions is played by the concept of an order parameter, that physical quantity that occupies a key place in the processes leading to the transformation. In the theory of percolation clusters, the order parameter is the power of an infinite cluster - the probability that a lattice node will belong to an infinite cluster. The critical behavior of this quantity at $p \to p_c$, $p > p_c$ is determined by the dependence

$$P_\infty = (p - p_c)^\beta, \tag{1}$$

where β is one of the so-called critical indicators (scaling indices - in terms of percolation theory) [1, 4]. The value β determines the critical behavior of the power of an infinite cluster $P_\infty$. In percolation theory, probability (1) is also called the percolation probability. It serves as the main characteristic of the percolation system. The percolation probability can be used to express properties of physical systems that depend on the topology of large clusters, such as spontaneous magnetization or conductivity. Such quantities as the average number of nodes of the final cluster, the correlation length ξ, the characteristic spatial scale of the cluster at $p<p_c$, and at $p>p_c$ - the characteristic size of the voids in it are also determined.

The value of $P_\infty$ in a nuclear reactor is proportional to the deviation $n - n_c$, where $n$ is the neutron density, $n_c$ is the critical value of $n$. Let us denote $\tau=(p-p_c)/p_c=k_{ef}-1$. The average number of nodes of a finite cluster $s$ (this value is similar to the generalized susceptibility in statistical physics and the fluctuation theory of phase transitions [8, 9]) at $\tau \to 0$ behaves as

$$s \sim |\tau|^{-\gamma}, \quad \tau \ll 1, \tag{2}$$

where $\gamma$ is the critical index for $s$. The correlation length ξ, the characteristic spatial scale of the cluster at $p<p_c$, and at $p>p_c$ - the characteristic size of voids in it, near the critical point the average size of the critical cluster at $\tau \to 0$ behaves as

$$\xi \sim |\tau|^{-\nu}, \tag{3}$$

where $\nu$ is the critical indicator of the value ξ (relative to $\tau$). For nuclear reactors, the influence of external influences (sources, control rods, feedback, size of the system, its boundaries, etc.) is significant. If we denote the external field $h$ (set in some effective way depending on the type of influence), then at $h \neq 0$ we can introduce another critical indicator $\delta$

$$P_\infty(\tau=0, h) = h^{1/\delta}. \tag{4}$$

Since $P_\infty \sim n - n_c$, then external fields, which are sources and sinks of neutrons in nuclear reactors (their physical nature is due to the influence of feedback, control effects, the influence of boundaries, delayed neutrons, etc.), shift the critical point, playing the role of reactivity additives. The dependence of reactivity on such additives is generally nonlinear, and changes in reactivity are not additive. The value of $p_c=1/\bar{\nu}$ shifts and fluctuations $\bar{\nu}$ - the number of neutrons emitted during the decay of the nucleus. The analogue of thermodynamic heat capacity $c_T$ behaves at $\tau \to 0$ as

$$c_T \sim |\tau|^{-\alpha}, \tag{5}$$

where $\alpha$ is the corresponding critical indicator. Formulas (2) and (3) are known in the theory of nuclear reactors, although they were obtained there in a different way. Thus, expression (2) is the neutron multiplication equation:

$$N = (1 - k_{эф})^{-1}, \tag{6}$$

those. critical index $\gamma=1$, equation (3) - critical size equation

$$R_{эф}=\pi M(k_{ef}-1)^{-1/2}, \tag{7}$$

where $R_{ef}$ is the effective size, a geometric parameter, $M$ is the neutron migration length. In this case, the indicator $\nu = 1/2$. Expression (4) reflects the influence of factors such as control rods and boric acid concentration on the value of the order parameter at the critical point. It is not easy to give an unambiguous interpretation of expression (5). But the use of percolation theory and the constructions of fractal theory makes it possible to write down a number of other relationships and consider, for example, dynamic critical indices, the dimension of the cluster skeleton, the spectral (fracton) dimension, etc.

Power dependences (as in expressions (1) - (3)) are generally characteristic of random variables with infinitely divisible distributions [18] and their subclass - stable distributions, the connection of which with fractal behavior has been strictly proven for the Wiener process, but which, apparently, are also associated with other random processes corresponding to fractal phenomena.

Critical exponents are related to each other by scaling relations

$$\alpha = 2 - d\nu = 2 - 2\beta - \gamma, \quad \delta = \gamma\beta + 1; \quad \eta = 2 - \gamma/\nu; \quad d\nu = 2\beta + \gamma, \tag{8}$$

where $d$ is the dimension of space, index $\eta$ determines the behavior of the correlation function $g(r)$ at large $r$, when $g(r) \sim r^{-d+2-\eta}$. The classical theory (its presentation for the case of polymer trees is given, for example, in [10]) allows us to obtain indices equal to $\gamma = 1, \beta = 1, \nu = 1/2, \delta = 2, \eta = 0, \alpha=1$. The validity of the classical indices for the reactor is explained by the fact that neutrons do not interact and the theory of a self-consistent field is valid. Critical exponents are universal, they do not depend on many rough properties of the chosen model, but are sensitive to various symmetry properties, the presence of long-range correlations, etc., characterizing a whole class of physical phenomena (in the asymptotic limit, near the transition point, where only the maximum term of the expansion is considered, for example by $\tau$ in expressions (2) and (3)).

The characteristic largest size of the finite clusters is determined by the exponent $\Delta$, for which the hyperscaling relation $\Delta = d\nu - \beta$ is valid. At $\tau \neq 0$

$$n_s(\tau) = f(s)\exp(-As|\tau|^\Delta), \tag{9}$$

where $n_s$ is the fraction of fission chains, clusters consisting of $s$ neutrons; $f(s)$ is some function that grows no faster than the power $s$. Clusters with $s_c \sim |\tau|^{-\Delta}$ are called critical. The probability that clusters will have larger sizes decreases exponentially. The characteristic spatial scale of the critical cluster is equal to the correlation length.

The behavior of the system is determined by the relationship between two spatial scales: the minimum length $a_0$ (a value on the order of the neutron free path $\lambda$) and the correlation length $\xi$. At $\xi >> a_0$ there is a region of intermediate asymptosis $a_0 << r << \xi$. In this region, all characteristics of clusters (measurements on a scale smaller than $\xi$) are similar to their characteristics at the most critical point, when $\tau = 0, \xi = \infty$. Their properties in this area are characterized by self-similarity (scale invariance). The reason for the similarity of critical phenomena is the similarity (self-similarity) of geometric objects. A characteristic of such self-similarity is the fractal dimension. At scales smaller than $a_0$, there is no self-similarity. There are various ways to determine the fractal dimension. For example, in the relations from [5, 15], the total number of particles in the system $N$ is related to the linear size of the system $r$ by the relation

$$N < r^D, \tag{10}$$

where $D$ is the self-similarity dimension, the fractal critical dimension. In [15], a relation was written for the length of a broken curve of the form

$$L = a_0 (R/a_0)^D,$$

where $a_0$ is the value of the scale used; $R$ – distance in a straight line; $D$ – fractal dimension. Near the critical point, the system can be considered as fractal self-similar, on scales $a_0 << r << \xi$, and as homogeneous on large scales. You can enter density $\rho_N$ type

$$\rho_N = \frac{N}{r^d} = \begin{cases} r^{D-d}, & r < \xi \\ const, & r > \xi \end{cases}. \tag{11}$$

The value $\xi$ is the scale at which the density becomes constant (in NR this is the region in which nuclear fissions do not occur, but there is a uniform flux of neutrons escaping from the region with fissions, for example, a reflector). Any intensive quantity behaves in the same way. Fractal behavior significantly affects the operation of nuclear reactors in the critical region. Thus, fractal characteristics determine the speed of propagation of disturbances in the supercritical regime, at $p > p_c$. The chemical distance $R_c$ between nodes $i$ and $j$ is defined as the minimum number of steps in which one can get from $i$ to $j$, passing only through accessible nodes. Having defined the ball $B_c$ as a set of nodes for which $R_c \leq n$, we can define the chemical dimension as an indicator $D_c$ such that the number of $N_B$ nodes belonging to $B_c$ increases with $n$, as

$$N_B \sim n^{D_c}. \tag{12}$$

The value $D_c = D/D_R$ is the ratio of two dimensions - the cluster dimension $D$ and the dimension $D_R$ of the curve, the length of which gives us the chemical distance. Another name for this dimension is the connectivity dimension. The effective speed of propagation of the disturbance front at $p > p_c$, at local supercriticality is equal to $v \sim \tau^{(D_R-1)\nu}$; the indicator $\psi = (D_R - 1)\nu$ effective speed determines the magnitude of the critical deceleration as $p$ approaches $p_c$ from above. In classical theory $\psi = 0.5$, and $v \sim \tau^{0.5}$. This characteristic is important for assessing various types of time characteristics of a critical reactor and, ultimately, for the safety of nuclear power plants. If local supercriticality occurs in a nuclear reactor, then the speed of its propagation must be assessed based on fractal patterns. Many important properties of the kinetics of processes in nuclear radiation can be understood in more detail by considering the dynamics of processes on the Bethe lattice (taking into account delayed neutron sublattices); this applies to both lattice diffusion and other dynamic phenomena. From the fact that $\psi=0.5$, we find: $D_R = 2$, $D_c = 2$ and $N_B \sim n^2$.

For the Bethe lattice and Cayley trees $d = 6$, $D = 4$. Classical values of the critical exponents are valid for $d \geq 4$ [9]. Note that fractals can have an entire fractal dimension greater than the dimension of the space in which they exist; this is true, for example, for infinite clusters. Cayley trees are constructed in ultrametric hierarchical space. For a three-dimensional Euclidean embedding space in a cubic lattice, the value of the fractal dimension corresponds to the branching measure. This paper examines approaches to describing fission processes in nuclear reactors associated with percolation laws, with fractal dimension $D = 4$ and Laplacian fractals with $D = 2.4$. It is also possible to use other models, for example random trees [17].

Expression (4) makes it possible to evaluate the influence of control parameters (using control rods, boric acid concentration, etc.) at the most critical point, with reactivity $\rho$ equal to zero. The classical value in the theory of a self-consistent field of the critical index is $\delta=2$. From expression (6), differentiating it with respect to $\tau \sim \rho$, we obtain: $\dfrac{dN}{d\rho} \sim -|p-p_c|^{-2}$, and the power coefficient of reactivity near the critical point is equal to $\alpha_N \sim \dfrac{d\rho}{dN} \sim -|p-p_c|^2 \sim -|k-1|^2$.

An expression is also written for the behavior of the correlation function $g(r)$ at large $r$, when $g(r) \sim r^{-d+2-\eta}$, the classical value of the index $\eta = 0$, and for $d = 6$, $D = 4$:

$$g(r) \sim r^{-d+2} \sim r^{-4} \sim l^{-2}.$$

Note that in the Cayley tree model, relations of the form (10) involve not the Euclidean distance $r$, but the chemical distance $l$ between nodes. Thus, the chemical distance between the central node and an arbitrary node belonging to the $l$-th shell is equal to $l$ [26]. The last relation $l^{-2}$ corresponds to the dependence $l^{-(d-D)}$. The size of the critical cluster is determined by the relation $s_c \sim |\tau|^{-\Delta}$, where the value $\Delta = 2$. For $d=3$, in a cubic lattice, $g(r) \sim r^{-d+2} \sim r^{-1}$. The total number of particles in the system $N$ is related to the chemical distance in the system $l$ by the relation $N \sim l^{D_c}$. The fraction of clusters with $s$ nodes (the fraction of fission chains of s neutrons) in accordance with expression (9) and with the classical value of the index $\Delta = 2$ is equal to $n_s(\tau) = f(s)\exp(-As|\tau|^2)$, $n_s(p_s) = s^{-5/2}$. The last expression shows that at the most critical point the fraction of fission chains of $s$ neutrons is related to the value of $s$ not by an exponential, but by a power-law dependence.

For neutron processes in a nuclear reactor, the most important characteristics are the percolation probability, which is interpreted as the probability of a self-sustaining chain reaction, and the percolation threshold value, which is proportional to the neutron multiplication factor. In [20], a recurrence relation was obtained for the probability of percolation from the root vertex, the probability that a connected component of the configuration containing the root vertex (some starting point of the appearance of the first neutron in the system, which generated a chain reaction), reaches the opposite edges of the system. Conventionally, mathematically, the size of the system and the connected component tends to infinity, although real systems are finite. It was noted in [21] that the size of the Bethe lattice is proportional to $lnN$, where $N=n$. The value $n$ in our problem is interpreted as the number of generations of neutrons in a chain reaction.

### 3. Models of irreversible growth. Dimensions

The fractal dimension of tree-like structures $D$ in a reactor with thermal neutrons and uranium as fuel (for example, VVER) is equal to the average number of secondary neutrons in one fission event, $D = 2.4$. This follows from [5, Ch. 16], where it is noted that for trees with infinitely thin trunks, the fractal dimension $D$ serves as a measure of branching. However, in this work we also used the Cayley tree model, for which the fractal dimension is $D = 4$. The difference in dimensions is caused by different embedding spaces. Another significant reason for the difference in fractal dimensions is the multifractality of the processes of fission and neutron transfer in nuclear reactors.

In [20, 21], fractal structures that simulate electrical breakdown are considered. The same kind of phenomena are observed in the processes of crack formation. The equations describing these processes are compared with the diffusion model of neutron transfer in the reactor. The study is carried out on a cubic lattice in a three-dimensional embedding space. The fractal dimension $D$ for this case is close to the branching measure of 2.4 for uranium nuclei in the reactor. The human circulatory system has almost the same dimension: 2.4 – 2.6.

The average density of particles in a cluster behaves as it moves away from the center, as in relation (11). It can be expressed by the formula [7]

$$\rho_N = \rho_{N0}(\frac{r_0}{r})^{\alpha_1}\Phi(\frac{R-r}{\sqrt{2}\Delta r}), \tag{13}$$

where $\alpha_1 = d - D$, $d$ – dimension of space, $\alpha_1 = d - D = 0.6$ for $d = 3$, $D = 2.4$; $\Delta r$ is the depth of penetration of a particle into a cluster of size $R$; $r$ - distance from the cluster center; $R >> R - r_0 >> \Delta r$; $\Phi(x)$ - probability integral; $r_0$ is a certain minimum radius (such as the mean free path of a neutron). For the total number of particles in the cluster we find (for $\Delta r << R$)

$$N = \int_0^R 4\pi r^2 \rho(r)dr = 4\pi r_0^{\alpha_1}\rho_0\frac{R^D}{D}[1-(\frac{2}{\pi})^{1/2}D\frac{\Delta r}{R}], \quad D = 2.4. \tag{14}$$

It is possible to estimate the dependence of the particle penetration depth $\Delta r$ when connecting to a cluster from the task parameters. The expressions obtained in [20] for the average distance between nearest branches may also be of some interest in reactor theory

$$\langle\lambda(r)\rangle = r^{(\frac{d-D}{d-1})}, \tag{15}$$

and other parameters. If $\gamma(r)$ is the number of lines of neutron trajectories of one chain intersecting a circle of a given radius; $\gamma_\pm(r)$ is the radial density of branching points (fissile nuclei) (+) and free ends (-) (neutron absorption points), and $L(r)$ is the total length of the neutron trajectory lines of one chain in a circle of radius $r$, then the following are satisfied similarity relations [21]

$$L(r) \sim r^D; \quad \gamma(r) \sim r^{D-1}; \quad \gamma_\pm(r) \sim r^{D-2}. \tag{19}$$

In addition to those listed, you can specify other fractal characteristics that can prove to be effective for solving various problems of operation and studying the behavior of nuclear reactors. Relations (15) and (19) are written for $d = 3$, $D = 2.4$ in the form

$$\langle\lambda(r)\rangle \sim r^{0.3}; \quad L(r) \sim r^{2.4}; \quad \gamma(r) \sim r^{1.4}; \quad \gamma_\pm(r) \sim r^{0.4}.$$

In [17], such an internal indicator of a cluster is used as the internal dimension of its skeleton $d_l^S$. The cluster skeleton is defined as the set of all shortest paths connecting the cluster nodes with the surrounding shell, located at a chemical distance $L << l$ from the region under consideration. The skeletons of percolation clusters in a space of any number of dimensions at a critical concentration are linear in the sense of the chemical distance, i.e. $d_l^S = 1$. General relationships between dynamic indicators describing diffusion and statistical indicators describing the structure of the tree are obtained:

$$d_w^l = 2 + D_c - d_l^S; \quad d_w = D(2 + D_c - d_l^S)/D_c; \quad \bar{d} = 2D_c/(2 + D_c - d_l^S),$$

where $d_w$ is the diffusion index; $d_w^l$ – chemical diffusion indicator; $\bar{d} = 2D/d_w$ is the fracton (or spectral) dimension characterizing the connectivity of an object, the density of states and its spectral properties [17]. When $d_l^S = 1$,

$$d_w^l = 1 + D_c; \quad d_w = D(1 + D_c); \quad \bar{d} = 2D_c/(1 + D_c).$$

For Cayley trees $D_c = 2$, $d_w = 6$, $d_w^l = 3$, $\bar{d} = 4/3$. These are not all the results of fractal and percolation approaches to describing the internal and dynamic properties of neutron clusters that form tree-like structures of fission chains in nuclear reactors. Thus, in [1, 2] the dimension of the shell, the boundary of a connected cluster, is defined as $D_h = (1 + \nu)/\nu$, equal to 3 for percolation on Cayley trees. In [22], the dependence of the cluster perimeter $P$ on the number of nodes in it $s$ is given by the relation

$$P = [s(1 - p_c)/p_c] + As^\sigma,$$

where $A$ is a constant, $\sigma = 1/\nu D$, equal to 1/2 for percolation on Cayley trees. In [22], the fractal dimension of the cluster perimeter $d_G$ (equal to 2 for percolation on Cayley trees), obtained in the butterfly walk (cluster growth) model, was considered, and a new relation was proposed for the chemical dimension $D_c = D/(D - d_G)$.

## 4. Conclusion

The importance of the relations of the percolation theory for neutron processes in a reactor is already evident from the fact that they allow us to immediately obtain the neutron multiplication equation and the equation for the critical size of the reactor (expressions (6) and (7), interpreting the general relations of the percolation theory (2) and (3)). Expressions (6) and (7) obtained directly from the relations of percolation theory indicate the effectiveness of this approach in the theory of neutron processes in a reactor. The relation for the speed of propagation of a disturbance at local supercriticality should prove useful. Many other expressions of percolation theory applied to reactors may also be of interest. This is apparently due to the fact that percolation is a critical process that presupposes the existence of a threshold, a critical point. At the threshold, flow occurs along a fractal set, the geometry of which is determined by criticality. The geometric characteristics of a fractal are independent of the microscopic properties of the medium. Below the critical point, kinetic processes are limited to a finite region of phase space, scattering, absorption and other neutron processes. At the critical point, the fractal set, which is formed when the free energy of the statistical ensemble decreases, becomes decisive. The behavior of the system under slow forcing influences on it tends to self-organized criticality - a singular nonequilibrium (quasi)stationary state [23, 24]. Stationary nonequilibrium states on fractal structures are chaotic, turbulent in nature. In [25 - 26], the Lorentz model is used to study them.

The kinetics and processes of transfer in fractal reactor structures require a separate detailed study [4, 11, 12, 26]. In the region of the critical point, long-range correlation effects appear, manifested in the non-Gaussian behavior of kinetic processes, determined by the topological invariants of self-similar fractal sets. Transfer processes at the percolation threshold are discussed in [11 - 12]. Fractional derivative equations are used that take into account the effects of memory, nonlocality and intermittency.

A deeper analysis of the hierarchical structure of neutron trajectories in nuclear reactors, based on the results of [26], shows that the behavior of the probabilities of a chain reaction occurrence is determined by the probabilities $c = p = \lambda_f/(\lambda_f + \lambda_c)$, the degree of criticality, and proximity to the critical point. Depending on this proximity, three (more precisely, four, if we highlight the critical point itself) main modes of behavior are distinguished: subcritical and supercritical (in them the laws of behavior (4) and (10) differ only in sign), critical (11), (14), and critical point (12), (13). In the traditional theory of nuclear reactors, only subcritical and supercritical regimes and the critical point are studied, although in the general theory of phase transitions a critical region is necessarily present. This is due to the fact that neutrons do not interact; the values of the classical critical indices are valid for them (as for a self-consistent field) [31]. In the stationary operating state of reactors there are many neutrons, their number can be considered infinitely large. In this case, the critical region contracts to a critical point. Note that in this case the equation for *P* in the continuum limit can be solved exactly. But the integrals are complex, and it is difficult to express the function *P* explicitly.

The critical region itself, as shown in [32], has a complex three-member structure. In [32], three modes of critical behavior of nuclear reactors were discovered, depending on the sign of control actions and feedbacks, the boundaries of these modes were found, and it was shown that in the region of the critical point the time behavior is power-law. Time is proportional to the number of generations, and this behavior is characteristic of self-similar irregular trees [26]. At the most critical point, the total number of neutrons is proportional to time, which corresponds to a degenerate tree. Thus, neutron trajectories vary depending on the probability c and the multiplication factor. In the subcritical (and supercritical) region the movement occurs along regular trees, in the critical region along self-similar irregular trees, at the critical point along a degenerate tree. Above the critical point, but in the critical region - again using self-similar irregular trees. In the supercritical region - again using regular trees.